# Nanostructured $La_{0.5}Ba_{0.5}CoO_3$ as cathode for solid oxide fuel cells

Augusto E. Mejía Gómez[1,2], Diego G. Lamas[3,4], Ana Gabriela Leyva[1,2,3], Joaquín Sacanell[1,2,*]

[1] Depto. de Física de la Materia Condensada, Gerencia de Investigación y Aplicaciones, Centro Atómico Constituyentes, CNEA, Av. Gral. Paz 1499, San Martín (1650), Buenos Aires, Argentina.
[2] INN, CNEA-CONICET, Av. Gral. Paz 1499, San Martín (1650), Buenos Aires, Argentina.
[3] Laboratorio de Cristalografía Aplicada, Escuela de Ciencia y Tecnología, Martin de Irigoyen 3100, Edificio Tornavía, Campus Miguelete, UNSAM, San Martín (1650), Buenos Aires, Argentina.
[4] CONICET, Av. Godoy Cruz 2290, CABA, Argentina.

**Abstract**

A simple method has been used to synthesize nanostructured $La_{0.5}Ba_{0.5}CoO_3$ (LBCO) powders, by confining chemical precursors into the pores of polycarbonate filters. The proposed method allows us to obtain powders formed by crystallites of different sizes, it is scalable and does not involve the use of sophisticated deposition techniques.

The area specific polarization resistance of symmetrical cells was studied to analyze the electrochemical behavior of the LBCO nanostructures as cathodes for Solid-Oxide Fuel Cells.

We show that the performance is improved by reducing the size of the crystallites, obtaining area specific resistance values of 0.2 $\Omega cm^2$ at 700ºC, comparable with newly developed cathodes using novel deposition techniques.

**Introduction**

Solid-Oxide Fuel Cells (SOFC) are electrochemical devices for energy production. Their high efficiency, long-term stability, fuel flexibility and low emissions, position them amongst the best potential candidates for clean sources of energy. Their largest disadvantage is the high temperature needed for operation (~1000ºC), which results in longer start-up times, degradation of the constituent materials, mechanical and chemical compatibility problems, etc. For those reasons, significant effort is devoted to develop materials able to operate in the so-called intermediate temperature (IT) range (500-800ºC). LaSr cobaltites are, in this sense, suitable to face the problem, since they have high electronic and oxide ion conductivity in the IT range, both of which are prerequisites for the cathodic process to occur. However, La;Sr cobaltites can be subject of degradation thus negatively affecting long term performance [1,2] and therefore, several related compounds are considered as alternatives. Among them, $La_{0.5}Ba_{0.5}CoO_3$ (LBCO) has been shown to display excellent cathodic properties [3,4,5,6,7].

---

[*] Author to whom correspondence should be addressed. Electronic mail: sacanell@tandar.cnea.gov.ar

Besides the exploration among different compounds, many efforts have been devoted to study the influence of microstructure on the electrochemical properties of the materials used as SOFC components. The use of nanostructures is known to be beneficial for the electrochemical performance of fuel cell components, as compared with microstructured materials, a fact that has been shown in several compounds typically used for cathodes [8, 9]. However, nanostructures of LBCO are difficult to synthesize with the typical techniques used in the literature for this compound, mainly due to the high temperature needed for the synthesis.

An interesting breakthrough has recently been achieved using pulsed laser deposition (PLD), obtaining materials with very fast exchange kinetics in $LnBaCo_2O_{5+d}$-based cathodes (Ln=Pr, La) [10,11]. Although promising, those works deal with a mostly "restricted-to-lab" technique, not easily scalable for technological applications.

In this paper we developed nanostructured $La_{0.5}Ba_{0.5}CoO_3$, using a method that has been shown to improve the cathodic properties of other similar compounds [2,8,12,13]. By this method we were able to obtain samples with crystallites of different sizes in the nanometric scale. In contrast with other deposition procedures, the method proposed here is very simple, allows us to obtain nanostructured samples, it is easily scalable to perform electrodes of different sizes, it does not involve the use of expensive techniques and gives rise to a comparable performance. Also, the performance can be further improved by testing alternative attaching procedures for cathode-electrolyte sintering.

**Experimental**

The synthesis was performed following the pore wetting of porous polycarbonate templates with an aqueous solution of the appropriate reagents in stoichiometric proportions. Then several thermal treatments were conducted in order to obtain the desired compound. A 0.25M stoichiometric solution of $La(NO_3)_3.6H_2O$, $Ba(NO_3)_2$ and $Co(NO_3)_2.6H_2O$ was prepared by dissolution of analytical reagents in pure water. To avoid the cobalt oxide precipitation the solution was maintained at acidic pH. Templates of porous polycarbonate films were used as filters in an adequate system for syringe filtration. By this process the total volume of the pores is filled with the solution. Commercial porous polycarbonate films from Millipore[TM] were used as membrane filters with passing through holes of 200 and 800 nm diameter.

The reaction to obtain the desired compound proceeds by the denitration process of the confined precursor in a microwave oven. By adjusting the time and the energy applied to the sample it is possible to accomplish this reaction without producing damage to the polycarbonate film. The perovskite compound LBCO is finally obtained and the template is sacrificed during a thermal treatment in a standard furnace at the final temperature of 800ºC for 10 min. More details can also be found in references [12,14,15]. In the Supplementary Information we present SEM images of the powders obtained after the synthesis procedure (Figures S1 and S2).

The crystallite size variation was achieved by the avoidance of contact among intermediate products of the chemical reaction during the synthesis procedure, as will be discussed.

Samples were labeled as LBCO2 and LBCO8, for nanostructures obtained using templates with pores of 200 nm and 800 nm of diameter, respectively.

These nanomaterials were attached to $Ce_{0.8}Gd_{0.2}O_{1.9}$ (GDC) electrolytes to obtain symmetric LBCO/GDC/LBCO cells to study their cathodic performance. GDC electrolytes were prepared from Nextech Materials$^{TM}$ powders by uniaxial pressing at 200 MPa and sintering in air at 1350°C for 2 h, obtaining 0.5 mm thick discs with a relative density higher than 97%. The LBCO nanopowders were made into an ink for cathode deposition, using commercial ink vehicle (Nextech Materials$^{TM}$). We smeared this ink with a brush on the electrolyte to fabricate the symmetrical cells. Once painted, the samples were dried at 50°C in air for about 20 min. Cathode sintering was performed at 1050ºC for 1 hour, in order to obtain the best attaching conditions and an optimized performance. In the Supplementary Information we present SEM images of the surface and cross section of typical cathodes (Figures S3 and S4).

The cathodic properties were characterized by the area specific polarization resistance (ASR) in air at equilibrium conditions (zero bias), extracted from electrochemical Impedance Spectroscopy (EIS) measurements performed with a Gamry G750 potentiostat/galvanostat.

**Results**

In Figure 1 we show SEM micrographs of samples (a) LBCO2 and (b) LBCO8. All data were obtained after the powders were subjected to a thermal treatment of 1h at 1050ºC, i.e. the temperature used for cathode sintering.

In the case of LBCO2 we can see some "large" (~ 1 µm long) rod-like structures and a majority of more heterogeneous structures in the same length scale. In the case of LBCO8, almost all the sample displays those heterogeneous structures. In most of our previous experience with the present synthesis procedure in other compounds, the resulting sample displayed a rod/tubular structure [2,8,12,16]. The morphology of the present samples does not show that morphology in general, a fact that could be due to the comparatively less concentrated solution of the precursors that was used in the present work in order to avoid segregation of the reagents (0.25M as compared with 1M used in the previous cases).

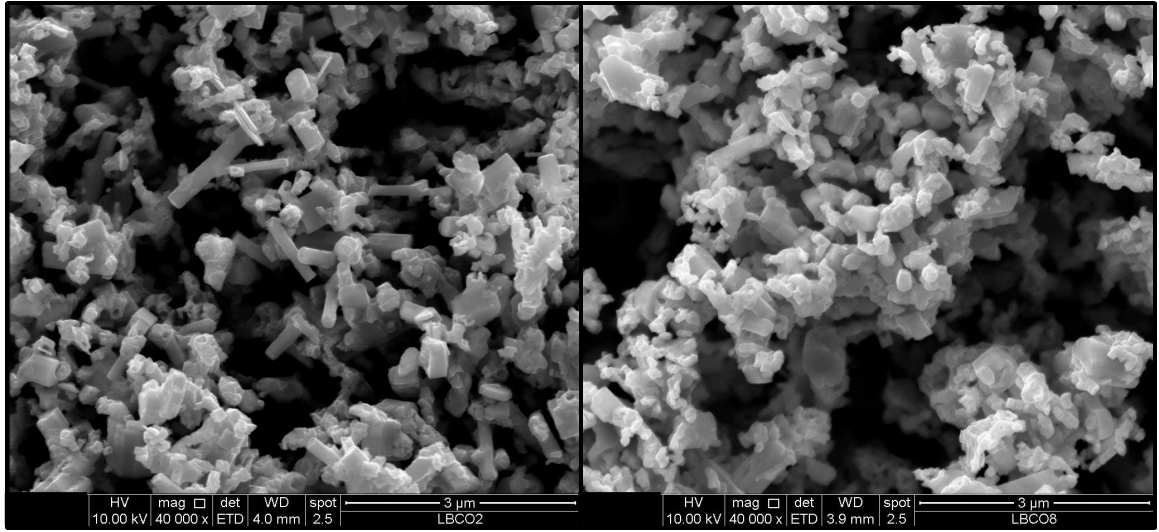

Figure 1: SEM micrographs of the (a) LBCO2 and (b) LBCO8 samples. Both cathodes were sintered at 1050ºC for 1 hour.

The XRD study of samples treated at 800°C showed that they were not single-phased, exhibiting a mixture of $LaCoO_{3-d}$ and $BaCoO_{3-d}$ phases (see Supplementary information). In contrast, it was found that samples treated at 1050ºC, the temperature used for cathode sintering, were single-phased. XRD data corresponding to LBCO2 and LBCO8 samples treated at 1050ºC for 1 h are presented in Figure 2 and the relevant crystallographic information is summarized in Table 1. The nanostructured character of the samples is evidenced in the broadening of the Bragg peaks. It can be observed a comparatively larger FWHM for the XRD data of LBCO8, thus suggesting a smaller crystallite size. Both samples display a cubic perovskite-like structure, as reported for powders on the micrometric scale [3]. This is particularly different than the results obtained for samples in which La was replaced by Gd and Sm [13]. In that case, the crystal structure was a layered perovskite, although the same method was used to obtain the samples.

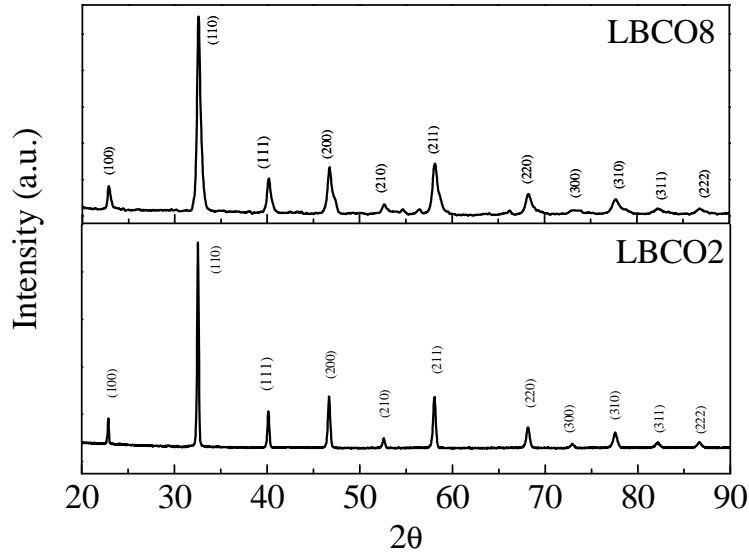

Figure 2: XRD patterns of LBCO2 and LBCO8 samples treated at 1050°C for 1 h.

Crystallite sizes, as obtained by the Scherrer equation, are 53 and 26 nm for samples LBCO2 and LBCO8, respectively, indicating that the structures shown in the micrographs of Figure 1 are composed by several crystals. We also observed a slight difference in the lattice parameters of the LBCO2 and LBCO8 samples, which is likely to be related to the particular synthesis conditions.

| Pore Size | Space group | a(Å) | b(Å) | c(Å) | $d_{XRD}$(nm) |
|---|---|---|---|---|---|
| 200 nm | Pm3m | 3,8811(9) | 3,8811(9) | 3,8811(9) | 53 |
| 800 nm | Pm3m | 3,8835(1) | 3,8835(1) | 3,8835(1) | 26 |

Table 1: Structural parameters obtained from the XRD data of Figure 2 for the powders of the LBCO2 and LBCO8 samples.

The lower value for the crystallite size of LBCO8 as compared with LBCO2, is likely to be due to the scarcity of contact points between the particles. As described in [14], in the present chemical route, the reaction to obtain the oxide starts inside the filled pores of a polycarbonate film. Then it proceeds by the partial denitration and dehydration of the confined precursors. After this point, intermediate oxides remain deposited on the surface of the template pores that still need to react to reach the final compound. The surface of the pores scales with their radius and therefore, a higher degree of dispersion is expected for templates with pores of larger diameters. Dispersed oxides result in lesser contact points among particles, which are the paths through which the particles grow. For that reason, a larger crystallite size is expected for LBCO2 as compared with LBCO8.

The electrochemical properties of our cathodes, measured in air at equilibrium conditions, are presented in Figure 3. The EIS data can be fitted with two components, as is evident from the data at 700ºC. Both cathodes show a similar dependence with temperature, with LBCO8 having the best performance. This can be inferred from the ASR values, defined as the base of the EIS arcs on the real axis (indicated in Figure 3 for the LBCO2 cathode at 700ºC, as an example), which measures the polarization resistance of the cathodic process.

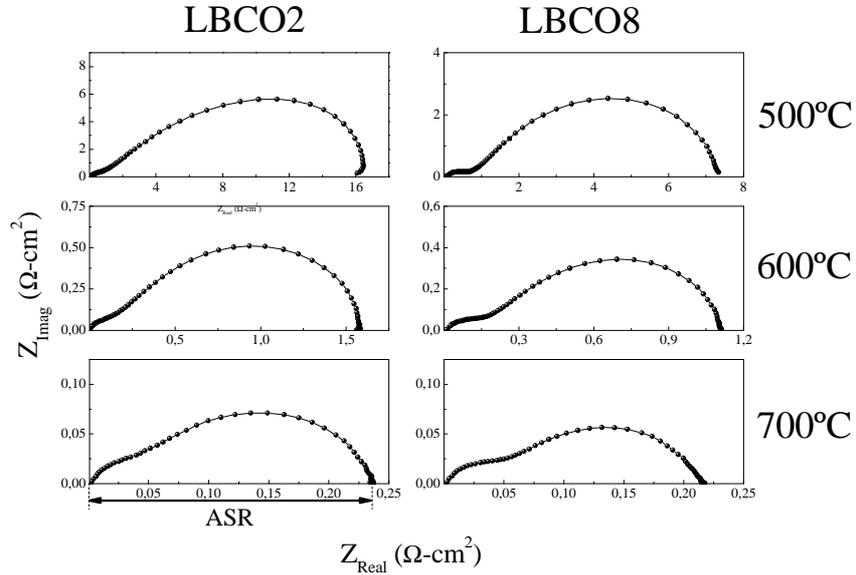

Figure 3: Electrochemical Impedance Spectroscopy results for the LBCO2 and LBCO8 samples cathodes. Measurements were performed in air at equilibrium conditions.

In Figure 4 the temperature dependence of the ASR of both cathodes is shown. We obtained activation energies of $E_a$ = 1.3 eV and 1.08 eV for the LBCO2 and the LBCO8 cathodes, respectively with no significant changes on varying p($O_2$) (not shown).

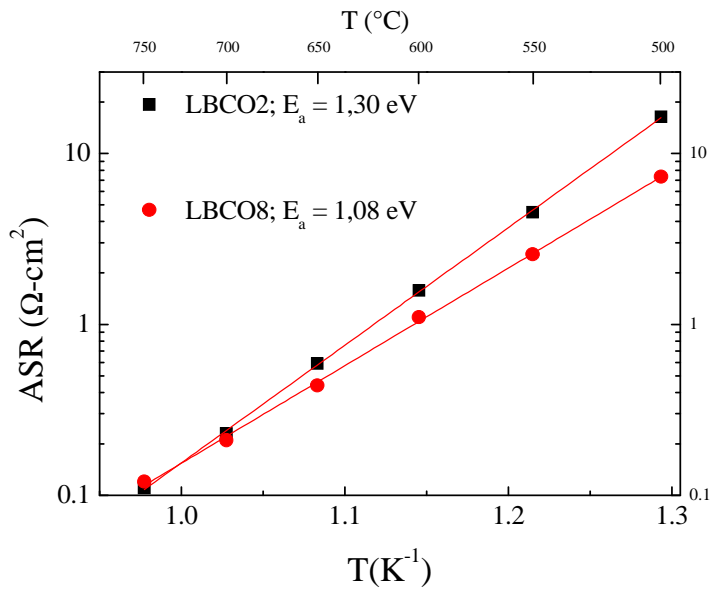

Figure 4: Arrhenius plot for the ASR values obtained from the electrochemical impedance diagrams.

The activation energies for the ASR are in agreement with previously reported data for the compound [17]. The Activation energy obtained for LBCO2 is similar to the ones obtained for equivalent nanostructures of $La_{0.6}Sr_{0.4}CoO_3$ with nearly the same crystallite size [2]. Thus, the lower activation energy obtained here for LBCO8 suggests that another mechanism is governing its overall reaction. As the crystallite size for LBCO8 is half of that of LBCO2, diffusion along grain boundaries will be more significant in LBCO8. That means that the contribution of grain boundary diffusion is relatively higher in the case of the LBCO8 cathode, as compared with LBCO2. As it has been shown that grain boundary diffusion is comparatively enhanced in comparison with bulk diffusion, this then is consistent with the overall better performance of the LBCO8 cathode.

In Figure 5, we show the EIS data at 700ºC obtained between $p(O_2) = 21\%$ (atmospheric pressure) and $p(O_2) = 5\%$, for (a) LBCO2 and (b) LBCO8. As previously shown in Figure 4, we can see that the LBCO8 cathode presents lower ASR than the LBCO2 one. We also see that on lowering the oxygen partial pressure, a low frequency process arises in LBCO8, which is virtually hindered for LBCO2.

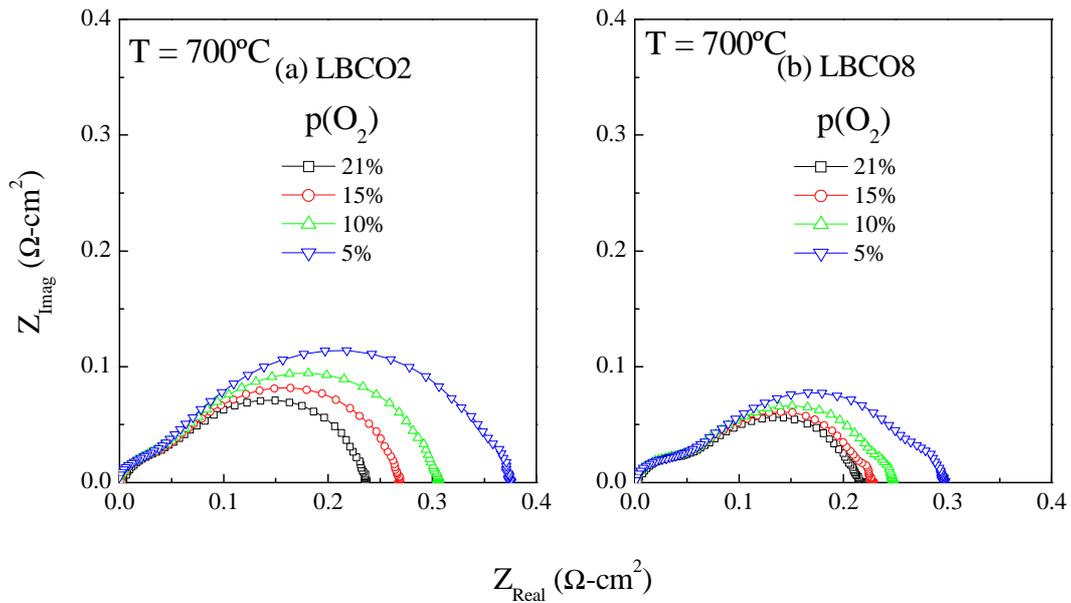

Figure 5: Electrochemical Impedance Spectroscopy results for the LBCO2 and LBCO8 samples cathodes for different oxygen partial pressures.

The low frequency process, is more evident in the case of the LBCO8 cathode (than in LBCO2) due to its comparatively small Warburg contribution. To gain insight onto the different contributions to the EIS spectra, we analyzed the obtained data with the equivalent circuit shown in Figure 6. For the dominant contribution at intermediate frequencies, we used a finite length Warburg element (W), as it is usual in MIEC materials as LBCO. In order to obtain an accurate fitting, we added an R-CPE parallel at high frequencies in all spectra and an additional one at low frequencies to account for the contribution observed at reduced p($O_2$). Figure 6 also shows the resistive parts of each of the elements that contribute to the equivalent circuit: $R_1$ is the high frequency contribution, $R_W$ is the resistive part of the Warburg element and $R_2$ the low frequency contribution.

To identify the dominant processes, we assumed the usual following dependence for the resistive parts of each component:

$$R_i \sim (p(O_2)^{n_i}) \qquad (1)$$

The exponent $n_i$ of eq. (1) gives information about the species involved in the corresponding process. In Table 2 we present the values of those exponents for each of the proposed contributions, obtained from the fitting of the data of Figure 6.

| Cathode | $n_1$ | $n_W$ | $n_2$ |
|---|---|---|---|
| LBCO2 | 0.17±0.09 | -0.32±0.02 | -1.1±0.2 |
| LBCO8 | 0.01±0.01 | -0.2±0.03 | -1.2±0.2 |

Table 2: Values of the $n_i$ index, as extracted from the fitting of the data of Figure 6 with eq. (1) for the LBCO2 and LBCO8 cathodes.

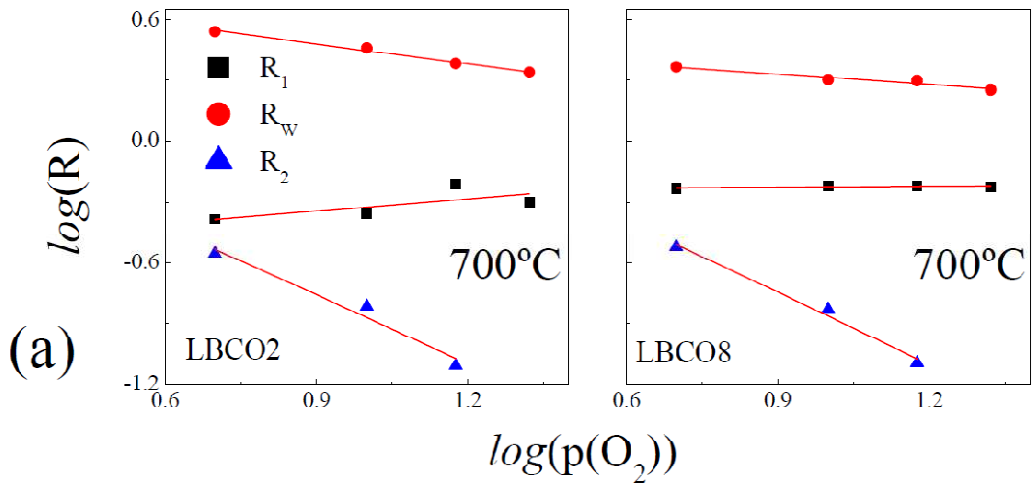

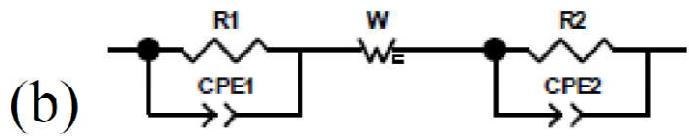

Figure 6: (a) Contributions to the polarization resistance as a function of the oxygen partial pressure at 700°C and (b) equivalent circuit used for fitting. The linear fittings observed in the log-log plot assume the dependence of equation (1).

We can see that in both cathodes, the Warburg contribution is clearly dominant, which is consistent with cathodes characterized by fast oxide ion transport. Also, we confirm that the low frequency contribution ($R_2 \| CPE_2$) can be ascribed to gas phase transport according to its $p(O_2)$ dependence ($n_2 \sim 1$) [18]. The high frequency contribution, with no clear, although slight, $p(O_2)$ dependence, is commonly ascribed to charge transfer across the cathode-electrolyte interface. Finally, the dependence of the polarization resistance, $R_p=R_1+R_W+R_2$ with $p(O_2)$, also follows a power law, with $n \sim 0.25$. This is consistent with an overall cathodic process limited by the oxygen exchange redox reaction on the electrode surface between the adsorbed oxygen and the bulk electrode material. Regarding more closely the dominant intermediate frequency contribution, we can see that the $p(O_2)$ dependence for the $R_W$ corresponding to LBCO2 is more significant than that of LBCO8, evidenced in the value of the $n_W$ index (see Table 2). An index of 0.23-0.28 has been usually attributed to surface ion conduction [19,20], while the adsorption of oxygen and the formation of oxygen ions is typically characterized by a larger value of the index (0.5 for the adsorption of atomic oxygen or dissociative adsorption [18,20], 3/8 for the formation of $O_{ad}^-$ [19], among others). Thus, the difference in the indexes from ~0.3 for LBCO2 to ~0.2 for LBCO8, can be related with a transition from an "adsorption/ion formation" dominating regime to a "surface diffusion" dominated one. Again, this is consistent with the smaller surface to volume ratio of LBCO2 as compared with LBCO8.

The results presented, allow us to depict the following image: nanostructured cathodes of LBCO display low ASR values, as needed for technological applications and moreover, the reduction of the crystallite size is beneficial in this aspect. As LBCO is a MIEC, this is undoubtedly related with two main facts: the increment of the reaction points for the oxygen reduction reaction and the enhancement of oxide ion conduction in the surface. Both features are directly related with the increment of the specific surface in nanostructured materials. The increment of the reaction points for the oxygen reduction reaction is straightforward on reducing the crystallite size. Also, an improved oxide ion conduction has been demonstrated in LBCO, mostly due to the difference in the ionic radii of La and Ba [3,4,21], which can be further enhanced (as shown in the present case) by the use of nanostructures in agreement with results obtained in other cathode materials [2,8,9,22].

**Conclusions**

In summary, we adapted a synthesis procedure to develop nanostructures of $La_{0.5}Ba_{0.5}CoO_3$, and we further used this material to produce symmetrical cells in order to characterize them as a potential cathode for SOFCs.

These cathodes display excellent performance, as can be inferred from the ASR values, positioning them as candidates for IT-SOFCs operating at temperatures lower than 800ºC.

By studying cathodes formed by powders with different crystallite sizes, we showed that a reduction of this parameter is beneficial for the cathodic performance.

The electrochemical impedance spectroscopy measurements also show that gas phase transport is optimized, due to the porosity of the samples, as its contribution can only be observed if the other components are largely reduced.

As a next step, we are performing studies of mechanical and chemical stability of the materials under operating conditions. Also, as the synthesis of nanostructures is feasible, we think that it should be interesting to study the possibility to perform a deposition of this materials on backbones of the electrolyte material, as successfully performed for other compounds.

**Acknowledgements**

Financial support from CONICET (PIP00038 and PIP00362) and ANPCyT (PICTs 1327, 1506 and 3411) is acknowledged. The authors thank Solange di Napoli for the critical reading and language revision of the manuscript.